\documentclass[twocolumn]{aastex631}

\usepackage[T1]{fontenc}

\usepackage{graphicx}
\usepackage{bm}

\usepackage{xcolor}
\usepackage{amsmath,amssymb,mathrsfs,gensymb}
\usepackage{yhmath}
\usepackage{breqn}

\newcommand{\new}[1]{#1}

\begin{document}

\title{It's not just a phase: oblique pulsations in magnetic red giants and other stochastic oscillators
}%

\author[0000-0002-1884-3992]{Nicholas Z. Rui}
\affiliation{TAPIR, California Institute of Technology, Pasadena, CA 91125, USA}%

\author[0000-0002-4544-0750]{Jim Fuller}
\affiliation{TAPIR, California Institute of Technology, Pasadena, CA 91125, USA}%

\author[0000-0001-7664-648X]{J. M. Joel Ong}
\affiliation{Institute for Astronomy, University of Hawai`i at Mānoa, 2680 Woodlawn Dr., Honolulu, HI 96822, USA}%
\affiliation{Hubble Fellow}

\date{\today}

\begin{abstract}
Magnetic fields play a significant role in stellar evolution.
In the last few years, asteroseismology has enabled the measurement of strong magnetic fields $10^4$--$10^6\,\mathrm{G}$ in the cores of dozens of red giants, and is the only known way to directly measure internal stellar magnetic fields.
However, current data are still interpreted assuming that these fields are too weak \new{or too axisymmetric} to affect the orientation of the pulsations (i.e., make the pulsations ``oblique''), rendering stronger field strengths beyond the reach of existing asteroseismic searches.
We show that, even when an oblique pulsator is also stochastic (such as in a red giant \new{with a strong non-axisymmetric magnetic field}), geometric effects will cause the signal to contain frequency components which remain in perfect relative phase with each other.
This perfect phase relationship persists even over timescales in which stochasticity erases absolute phase information.
This perfect relative coherence is a distinctive observational signature of oblique pulsation that does not require a model for mode frequencies to search for.
However, due to its dependence on phase, this effect will not be evident in the power spectral density alone, and phase information should be retained in order to detect it.
Coherence-based searches for oblique pulsations may pave the way to measurements of magnetic fields of currently inaccessible strengths in red giants, as well as some main-sequence and compact pulsators.
\end{abstract}


\section{Introduction} \label{sec:intro}

Although stellar magnetic fields are common, their formation, evolution, diversity, and role in angular momentum transport in stars and compact objects form a tangled web of open problems across astrophysics \citep{donati2009magnetic,ferrario2015magnetic,aerts2019angular}.
Our understanding of stellar magnetism is tethered to the uncertain strengths and structures of \textit{subsurface} magnetic fields, which are invisible to standard observational techniques.

Asteroseismology---the measurement and interpretation of stellar oscillations---exploits the translucency of stars to hydrodynamical waves to constrain internal stellar properties, such as stellar mixing processes, internal rotation profiles, and evolutionary states \citep{aerts2021probing}.
Magnetic fields with strengths $\simeq20$--$600\,\mathrm{kG}$ have recently been asteroseismically measured in the cores of several dozen lower red giant branch stars \citep{li2022magnetic,deheuvels2023strong,li2023internal,hatt2024asteroseismic}.
These measurements make use of the sensitivity of gravity-mode (g-mode) frequencies to the magnetic tension.
``Seismic magnetometry,'' which is still in its infancy, remains the only direct way to probe \textit{internal} stellar magnetic fields.

\new{When} incorporating magnetic fields, present data analyses assume the pulsations to be aligned with the rotation axis\new{.}
\new{This} occurs \new{either when the magnetic field is axisymmetric about the rotation axis, or otherwise is weak enough that its effects are subdominant to those of} the Coriolis and centrifugal forces.
In the opposite case, the magnetic field is strong enough to \textit{misalign} the pulsations from the rotation axis, i.e., make the pulsations oblique.
In this regime, individual oscillation modes appear as multiple, Doppler-shifted periodicities to the observer.
This breaks the one-to-one mapping between oscillation modes and frequency components in the light curve (hereafter ``periodicities''), and produces complicated pulsation spectra which can be difficult to interpret \citep{kurtz1982rapidly,shibahashi1993theory,dziembowski1996magnetic,bigot2002oblique,saio2004axisymmetric,loi2021topology}.
In some pulsators such as red giants, the oscillations are additionally \textit{stochastic}: each mode decoheres on a characteristic mode lifetime $\tau$ \citep[between tens of days to several years in red giants;][]{dupret2009theoretical,grosjean2014theoretical}.

In this Letter, we show that oblique, stochastically driven pulsators (such as red giants \new{with strong non-axisymmetric magnetic fields}) display some coherent properties which can be used to identify them in a general, model-independent way.
Because ordinary stochastic pulsations lack a mechanism for ``remembering'' phase information for times $\gg\tau$, this long-lived coherence is a smoking-gun signature of oblique pulsations.
Since this signature involves phase information, usual analyses based on the power spectral density (PSD) will be insensitive to it.

\section{Pulsation model} \label{pulsationmodel}

We construct a simple toy model which exhibits the essential behavior of stochastic, oblique pulsations.
The key observable is the intensity perturbation $\delta I(t)$, which has contributions from each oscillation mode:
\begin{equation} \label{deltaIfull}
    \delta I(t) = \Re\sum_j\delta I_j(t)\mathrm{,}
\end{equation}
where we have indexed the modes by $j$ and allowed the intensity $\delta I_j(t)$ of each individual mode to be complex.

The individual mode intensities are given by integrals of the surface flux perturbation $\delta F_j(t;\theta',\phi')$ over the visible disk of the star:
\begin{equation} \label{deltaI}
    \delta I_j(t) \propto \int^{2\pi}_0\int^1_0\delta F_j(t;\theta',\phi')W(\theta')\cos\theta'\,\mathrm{d}(\cos\theta')\,\mathrm{d}\phi'\mathrm{,}
\end{equation}
\new{where $W(\theta')$ is an arbitrary limb-darkening function on which the details of this analysis do not depend.}
The primed variables $(\theta',\phi')$ denote spherical coordinates in the \textit{inertial} (observer) frame, with the north pole ($\theta'=0$) fixed to the line-of-sight (direction pointing to the observer).
This is the frame in which integrals over the disk are most natural to compute.

Assuming surface flux perturbations trace scalar fluid perturbations \citep[e.g.,][]{gizon2003determining}, the flux perturbation due to a single mode can be decomposed into time- and angle-dependent factors:
\begin{equation} \label{fluxincorot}
    \delta F_j(t;\theta,\phi) \propto A_j(t)\psi_j(\theta,\phi)\mathrm{,}
\end{equation}
where the unprimed spherical coordinates $(\theta,\phi)$ are in the frame \textit{corotating} with the star, with the north pole ($\theta=0$) fixed to the rotation axis.
This is the frame in which the oscillation modes of the star are most natural to compute.

The angular dependence of a mode pattern can, in turn, be decomposed into spherical harmonics:
\begin{equation} \label{psispharm}
    \psi_j(\theta,\phi) = \sum_{\ell m}c_{j;\ell m}Y_{\ell m}(\theta,\phi)\mathrm{,}
\end{equation}
where $c_{j;\ell m}$ is the contribution of each spherical harmonic to the flux perturbation of mode $j$.
We show in Appendix \ref{wignerappendix} that, upon changing coordinates from the corotating frame to the observer's frame, these spherical harmonics transform as
\begin{equation} \label{wigner_specific}
    Y_{\ell m}(\theta,\phi) = \sum_{m'}e^{-im\Omega t}d^\ell_{m'm}(i)Y_{\ell m'}(\theta',\phi')\mathrm{,}
\end{equation}
where the Wigner matrix $d^\ell$ captures the effect of inclination on visibilities.

Combining Equations \ref{deltaI}, \ref{fluxincorot}, \ref{psispharm}, and \ref{wigner_specific} gives
\begin{equation} \label{subcombined}
    \delta I_j(t) \propto A_j(t)\sum_{\ell m} V_\ell \, c_{j;\ell m} \, d^\ell_{0m}(i)e^{-im\Omega t}\mathrm{,}
\end{equation}
where we have used the fact that spherical harmonics with $m\neq0$ have vanishing disk integrals, and defined
\begin{equation} \label{visibilities}
    V_\ell = \int^{2\pi}_0\int^{+1}_{-1}Y_{\ell0}(\theta',\phi')W(\theta')\cos\theta'\,\mathrm{d}(\cos\theta')\,\mathrm{d}\phi'
\end{equation}
to be the mode visibilities, which only depend on $\ell$.

The time dependence of a stochastically excited mode is well-described by a damped harmonic oscillator driven by noise \new{\citep[e.g.,][]{stello2004simulating}}.
As we show in Appendix \ref{stochasticappendix}, $A_j$ is well-modeled as
\begin{equation} \label{At_maintext}
    A_j(t) = \bar{A}_j(t)e^{i\sigma_jt}\mathrm{,}
\end{equation}
where $\sigma_j$ is the corotating mode frequency and the complex prefactor $\bar{A}_j(t)$ stays roughly constant for short times $t\ll\tau$ while varying randomly for long times $t\gtrsim\tau$.
We illustratively define $\bar{A}_j=H_je^{i\varphi_j}$, so that
\begin{equation}
    A_j(t) = H_j(t)e^{i(\sigma_jt + \varphi_j(t))}\mathrm{,}
\end{equation}
where the mode amplitude $H_j=H_j(t)$ and phase $\varphi_j=\varphi_j(t)$ are real-valued functions which, like $\bar{A}_j$, vary substantially only on timescales $\gg\tau$.

Our final expression for the total intensity perturbation follows from Equations \ref{deltaIfull} and \ref{subcombined}:
\begin{equation} \label{combinedIfull}
    \delta I(t) \propto \Re\sum_jA_j(t)\sum_{\ell m} V_\ell c_{j;\ell m}d^\ell_{0m}(i)e^{-im\Omega t}\mathrm{.}
\end{equation}

\new{A mode is ``rotationally aligned'' when its horizontal structure is well-described by a single spherical harmonic (i.e., $c_{j;\ell',m'}\approx\delta_{\ell\ell'}\delta_{mm'}$ for some $\ell$ and $m$).
In this special case, the sum over $\ell$ and $m$ in Equation \ref{combinedIfull} reduces to a single term, and the mode appears to an observer as a single sinusoidal signal.
Conversely, if more than one expansion coefficient $c_{j;\ell',m'}$ is nonzero, the mode is ``oblique,'' and the observed signal will be non-sinusoidal.
}

\subsection{Rotationally aligned pulsations}

\begin{figure*}
    \centering
    \includegraphics[width=\textwidth]{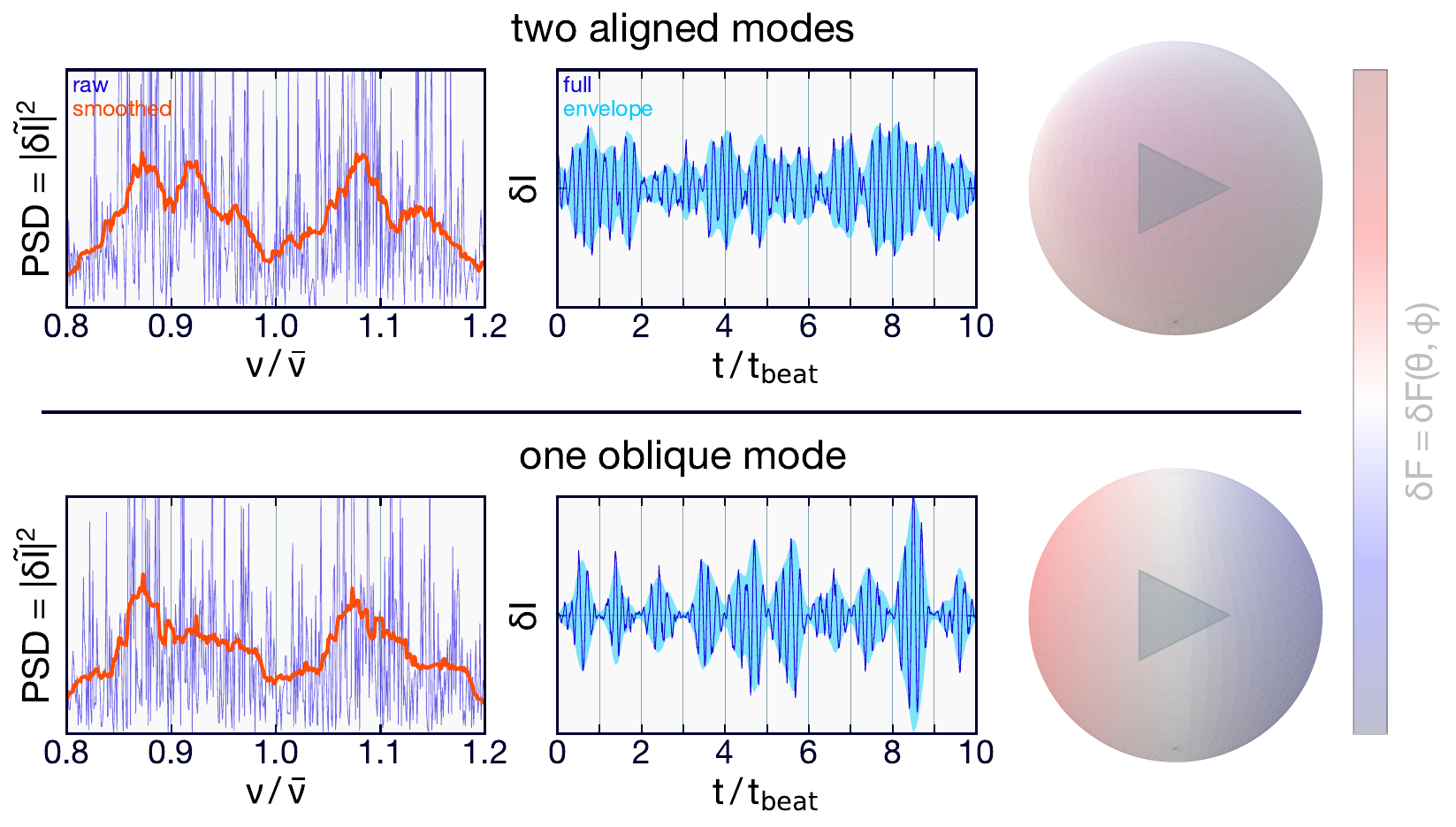}
    \caption{Comparison between two double-periodicity stochastic pulsations: two rotationally aligned modes (\textit{top}) and one oblique mode (\textit{bottom}).
    \textit{Left column}: Power spectral density (PSD).
    A boxcar-smoothed PSD is shown as a \textit{thick orange line}.
    \textit{Middle column}: Time series disk-integrated intensity, where the time is normalized to the beat period $t_{\mathrm{beat}}=2\pi/\delta\omega$.
    A Hilbert transform-based estimator for the beating envelope \citep{kress1989linear} is shown as a \textit{blue region} to guide the eye.
    \textit{Right column}: Animated map of the surface flux perturbation.
    These panels can be viewed in the animated version of this figure in the HTML version of the online article, or in a corresponding Zenodo upload at \new{\href{https://zenodo.org/records/15338766}{https://zenodo.org/records/15338766}}.}
    \label{fig:compare_modes_still}
\end{figure*}

It is common to assume that pulsations are \new{rotationally aligned} which, for g modes, occurs when the Coriolis effect supplies the strongest non-spherically symmetric restoring force.
Via Equations \ref{subcombined} and \ref{At_maintext}, the intensity perturbation due to such a mode is
\begin{equation} \label{rotationalign}
    \delta I_{\ell m}(t) \propto V_\ell\bar{A}_{\ell m}(t)d^\ell_{0m}(i)e^{i(\sigma_{\ell m}-m\Omega)t}\mathrm{.}
\end{equation}
Equation \ref{rotationalign} shows that the observer measures a Doppler-shifted mode frequency
\begin{equation} \label{dopplereffect}
    \omega_{\ell m} = \sigma_{\ell m} - m\Omega\mathrm{,}
\end{equation}
under the convention that positive-frequency modes with $m\Omega>0$ are retrograde.

Rotation also produces a Coriolis force which contributes to the restoration of fluid motions, causing an additional frequency shift
\begin{equation} \label{corioliseffect}
    \sigma_{\ell m} = \sigma^{(0)}_\ell + mC_\ell\Omega\mathrm{,}
\end{equation}
where $\sigma^{(0)}_\ell$ is the mode frequency in the absence of rotation, and the Ledoux coefficient $C_\ell\approx1/\ell(\ell+1)$ for high-radial order g modes \citep{ledoux1951nonradial}.

These effects taken together, the observer measures an apparent rotational frequency shift
\begin{equation} \label{rotationalshift}
    \delta\omega = \omega_{\ell m} - \sigma^{(0)}_\ell = -m(1-C_\ell)\Omega\mathrm{.}
\end{equation}
While $\delta\omega$ resembles a single shift proportional to $m$, it is actually caused by a combination of effects due to the Doppler shift (which is purely geometric) and the Coriolis force (which is physical).
These two very different effects become unnatural to group together when analyzing stochastic, oblique pulsators.

\subsection{Do two periodicities mean one mode or two?} \label{oneortwo}

When a pulsation is oblique (i.e., not rotationally aligned), a single oscillation mode is no longer a traveling wave of fixed $m$ around the rotation axis.
Instead, the morphology $\psi_j(\theta,\phi)$ of the mode has multiple components with different values of $m$ which Doppler shift into multiple periodicities via Equation \ref{dopplereffect}.
Because of this, a single oblique oscillation mode can therefore be misinterpreted as multiple (rotationally aligned) oscillation modes.
The so-called ``oblique pulsator model'' successfully describes the magnetically tilted pressure-mode (p-mode) pulsations of rapidly oscillating Ap stars \citep{dziembowski1996magnetic}.

Fortunately, if the modes are observed for long enough to resolve their stochasticity (i.e., for baselines $T\gtrsim\tau$), it is possible to tell whether two periodicities are caused by two distinct modes or a single mode which is oblique \textit{without} any model for their amplitudes and frequencies.
To illustrate this, we consider two toy scenarios which produce almost indistinguishable empirical PSDs peaked at two close frequencies ($\omega_1=\bar{\omega}-\delta\omega/2$ and $\omega_2=\bar{\omega}+\delta\omega/2$) despite their corresponding time series' obviously different properties (Figure \ref{fig:compare_modes_still}, in which $\omega_1=0.9\bar{\omega}$ and $\omega_2=1.1\bar{\omega}$).
The two periodicities can be resolved as long as $\delta\omega\gtrsim1/\tau$.

In the first scenario, we consider two rotationally aligned modes with frequencies $\sigma_1=\omega_1$ and $\sigma_2=\omega_2$ with quantum numbers $(\ell,m)=(1,+1)$ and $(1,-1)$, respectively.
The star is assumed to be non-rotating (or, for self-consistency, rotating at a negligible rate $\Omega\approx0$).
By Equation \ref{combinedIfull}, the intensity perturbation due to these two modes is
\begin{equation} \label{twomodes_raw}
    \begin{split}
        \delta I &\propto \Re\left[A_1 + A_2\right] \\
        &= \Re\left[e^{i\bar{\omega}t}\left(H_1e^{i(-\delta\omega t/2+\varphi_1)} - H_2e^{i(+\delta\omega t/2+\varphi_2)}\right)\right]\mathrm{,}
    \end{split}
\end{equation}
where we have used the fact that \new{$d^1_{0,-1}(i)=-d^1_{0,+1}(i)$} \citep{rose1995elementary} and omitted overall constant prefactors.
Defining $\bar{H} = (H_1+H_2)/2$, $\bar\varphi=(\varphi_1+\varphi_2)/2$, $\delta H=H_2-H_1$, and $\delta\varphi=\varphi_2-\varphi_1$,
Equation \ref{twomodes_raw} simplifies to
\begin{equation} \label{twomodes_final}
    \begin{split}
        \delta I &\propto 2\bar{H}\underbrace{\sin\left(\delta\omega t/2+\delta\varphi/2\right)}_{\mathrm{beat}}\underbrace{\sin\left(\bar{\omega}t+\bar{\varphi}\right)}_{\mathrm{carrier}} \\
        &- \delta H\underbrace{\cos\left(\delta\omega t/2+\delta\varphi/2\right)}_{\mathrm{beat}}\underbrace{\cos\left(\bar{\omega}t+\bar{\varphi}\right)}_{\mathrm{carrier}}\mathrm{.}
    \end{split}
\end{equation}

The intensity pattern consists of a high-frequency oscillation with a carrier frequency $\bar{\omega}$ modulated by a beating envelope with a lower frequency $\delta\omega$ (a factor of two arises because the envelope refers to the absolute value of the beat sinusoid).
On short timescales ($t\ll\tau$), the quantities $\bar{H}$, $\bar{\varphi}$, $\delta H$, and $\delta\varphi$ are all approximately constant, and the oscillation is roughly coherent.
However, for longer observation baselines $T\gtrsim\tau$, both the carrier and beat oscillations dephase, i.e., $\bar{\varphi}$ and $\delta\varphi$ vary randomly (\textit{top middle panel} of Figure \ref{fig:compare_modes_still}).

In the second scenario, we consider a single oblique mode with corotating frequency $\sigma=\bar{\omega}$ with two equal spherical harmonic components with $m=\pm 1$ (i.e., $c_{1,+1}=c_{1,-1}$).
Additionally, the star itself rotates with a rate $\Omega=\delta\omega/2$.
By Equation \ref{combinedIfull}, the resulting intensity perturbation due to this single mode is
\begin{equation} \label{oneobliquemode}
    \begin{split}
        \delta I &\propto \Re\left[A\left(e^{-i\delta\omega t/2} - e^{i\delta\omega t/2}\right)\right] \\
        &= \Re\left[-2iHe^{i(\bar{\omega}t+\varphi)}\sin\left(\delta\omega t/2\right)\right]\mathrm{.}
    \end{split}
\end{equation}
Upon taking the real part, Equation \ref{oneobliquemode} becomes
\begin{equation} \label{onemode_final}
    \delta I \propto 2H\underbrace{\sin\left(\delta\omega t/2\right)}_{\mathrm{beat}}\underbrace{\sin\left(\bar{\omega}t+\varphi\right)}_{\mathrm{carrier}}\mathrm{.}
\end{equation}

Similarly to the first scenario, the two periodicities generated by the oblique mode oscillate with a carrier frequency $\bar{\omega}$ modulated by an envelope with a beat frequency $\delta\omega$.
On short timescales ($t\ll\tau$), the intensity perturbation is indistinguishable from that of the first scenario, or, indeed, a purely coherent beat pattern.
However, while the overall amplitude $H$ and carrier phase offset $\varphi$ vary randomly as before, the beating envelope is perfectly coherent and \textit{never dephases}.
In the corresponding time series (\textit{bottom middle panel} of Figure \ref{fig:compare_modes_still}), the beating envelope vanishes on exact multiples of the beat period $t_{\mathrm{beat}}=2\pi/\delta\omega$.
Intuitively, the non-axisymmetric magnetic field misaligning the oblique pulsation serves as a ``clock hand'' which perfectly tracks the rotational phase.
In contrast, stars with no magnetic fields (or other non-axisymmetric features) have no mechanism by which to keep track of their absolute rotational phases over timescales $\gg\tau$.

Attempting to interpret the intensity perturbation generated by the single oblique mode as two separate rotationally aligned modes would imply the bizarre conclusion that the amplitude ratio $H_2/H_1$ and phase offset difference $\delta\varphi$ between the two stochastic modes are exactly constant in time.
Despite this, both the \textit{average} amplitude $H\rightarrow\bar{H}$ and phase offset $\varphi\rightarrow\bar{\varphi}$ would be observed to vary stochastically in the expected way.

By construction, the periodicities in both scenarios have identical frequencies and linewidths.
The two scenarios thus produce very similar-looking PSDs, each consisting of an envelope of two broad Lorentzians multiplied by noise (the \textit{left panels} of Figure \ref{fig:compare_modes_still}, cf. \citealt{cunha2020solar}).
Nevertheless, since the time-domain intensities $\delta I$ in the two scenarios are fundamentally different, the Fourier transforms $\delta\tilde{I}$ (which encode identical information) must also be different in some distinctive way.

The frequency-domain manifestation of oblique stochastic pulsation becomes apparent when comparing the Fourier transforms of Equations \ref{twomodes_final} and \ref{onemode_final}, which are
\begin{equation}
    \delta\tilde{I} \propto \frac{1}{2}\left[\tilde{\bar{A}}_1(\omega-\omega_1) - \tilde{\bar{A}}_2(\omega-\omega_2)\right] + \mathrm{sym.}
\end{equation}
for two aligned modes, and
\begin{equation}
    \delta\tilde{I} \propto \frac{1}{2}\left[\tilde{\bar{A}}(\omega-\omega_1) - \tilde{\bar{A}}(\omega-\omega_2)\right] + \mathrm{sym.}
\end{equation}
for one oblique mode, where ``$\mathrm{sym.}$'' denotes the frequency-flipped, complex conjugate of the first term (which arises from taking the real part).
In the case of two aligned modes, the Fourier profiles of the two periodicities ($\tilde{\bar{A}}_1$ and $\tilde{\bar{A}}_2$) are different, i.e., the noise multiplying the square-root Lorentzians in the Fourier transform are different from each other.
In contrast, in the case of one oblique mode, the Fourier profiles of the two periodicities are identical, i.e., the noise multiplying the square-root Lorentzians are the same across the two Fourier peaks.

Although frequency resolution, noise floor, and nonuniform sampling effects will cause non-ideal behavior, oblique pulsation will still generally produce spectral correlation which cannot be caused by separate, rotationally aligned, stochastic modes.
We note that spectral correlation describes correlation between frequency components, and is conceptually distinct from temporal correlation, which describes the time-domain correlations which are characteristic of colored (but stationary) noise.
\textit{Fourier peaks with identical noise profiles are smoking-gun signals of oblique pulsation} and, thus, the likely presence of a strong \new{non-axisymmetric} magnetic field.
Since the PSD ($|\tilde{I}^2|$) discards all of the phase information in $\delta\tilde{I}$, data analyses which start from the PSD are likely to overlook oblique pulsations.
\textit{It's not just a phase.}

\subsection{Magnetic red giants: a theoretical case study} \label{redgianttriplet}

\begin{figure}
    \centering
    \includegraphics[width=0.47\textwidth]{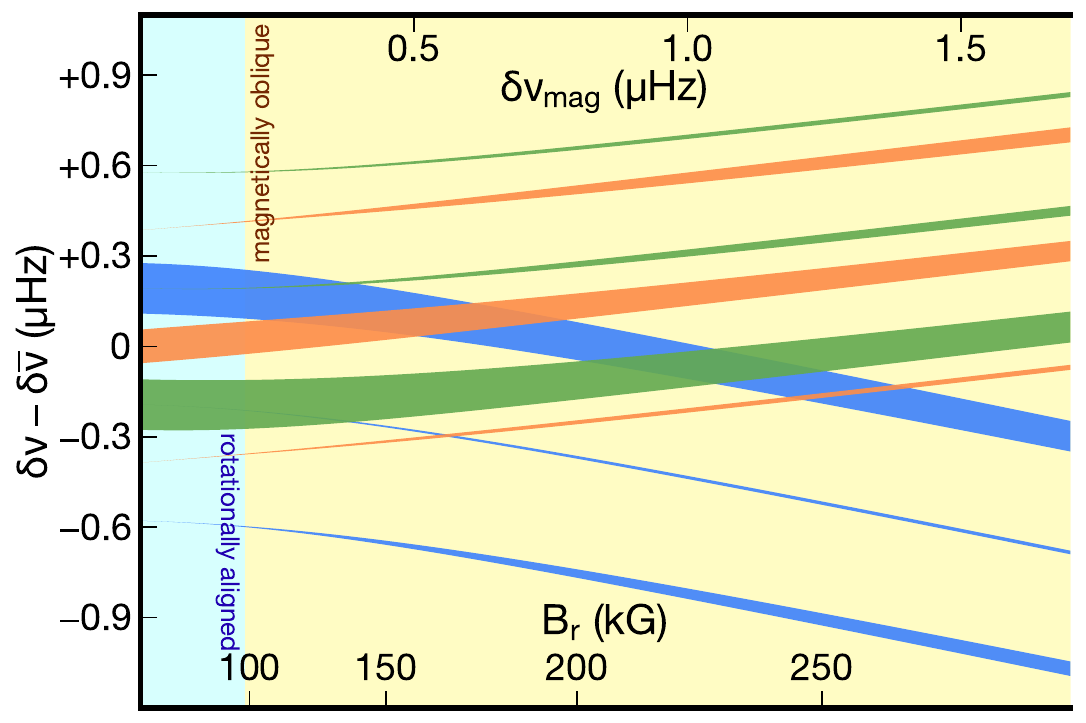}
    \caption{
    Linear g-mode frequency shifts $\delta\nu=\delta\omega/2\pi$ relative to the power-weighted average shift of the triplet $\delta\bar{\nu}$ \new{(defined in the main text)}, plotted against the strength of the magnetic perturbation $\delta\nu_{\mathrm{mag}}=\delta\omega_{\mathrm{mag}}/2\pi$.
    The magnetic field strengths corresponding to $\delta\nu_{\mathrm{mag}}$ are also shown for the $M=1.5M_\odot$, $R=5.2R_\odot$ red giant model described in Section \ref{redgianttriplet}.
    \new{For illustrative purposes here we have fixed $2\pi\sigma_0=180\,\mu\mathrm{Hz}$, $\delta\omega_{\mathrm{rot}}\approx1.2\,\mu\mathrm{Hz}$, $i=60\degree$, and $\beta=70\degree$.}
    Line colors indicate the underlying mode, and line thicknesses are proportional to the spectral power.
    Low field strengths leave the rotational triplet symmetrically split, moderate field strengths introduce asymmetry in the triplet, and higher field strengths induce oblique pulsations in which each mode individually appears as a triplet.}
    \label{fig:figure_magnetic_shifts}
\end{figure}

Seismically, red giants are especially rich: near-surface convection drives information-dense pulsations containing up to dozens of independent oscillation modes \citep{chaplin2013asteroseismology}.
In lower red giant branch ($R\lesssim10R_\odot$) and red clump stars, dipole ($\ell=1$) p modes are also well-coupled to g modes, which propagate exclusively in the stars' radiative cores.
Strong \textit{core} magnetic fields appreciably modify internal restorative forces, shifting mode frequencies in distinctive ways.
Measurement of these frequency shifts has recently enabled the precise extraction of core magnetic field strengths in dozens of red giants \citep{li2022magnetic,deheuvels2023strong,li2023internal,hatt2024asteroseismic}.

We briefly summarize known theoretical predictions of magnetic effects on red giant mixed modes \citep{li2022magnetic,mathis2023asymmetries,das2024unveiling}, deferring many details to Appendix \ref{magneticrg_appendix}.
Although red giant g modes can only be observed when coupled to p modes, the non-axisymmetric nature of the problem suggests that mixed-mode coupling may affect the spectrum in a complicated way.
Thus, for simplicity, we hereafter consider only pure g modes.

The seismic effects of rotation and magnetism are calculated by solving the eigenvalue problem
\begin{equation} \label{MR_eigenvalueproblem}
    \delta\sigma\,\mathbf{c}_\ell = (\mathbf{R}_\ell + \mathbf{M}_\ell)\,\mathbf{c}_\ell
\end{equation}
for the corotating frequency shifts $\delta\sigma$ and the spherical harmonic expansion coefficient vectors $\mathbf{c}_\ell$ (see Equation \ref{psispharm}).
The matrices $\mathbf{R}_\ell$ and $\mathbf{M}_\ell$ describe the effects of the Coriolis and Lorentz forces.

The $\ell=1$ Coriolis matrix $\mathbf{R}_{\ell=1}$ is given by
\begin{equation}
    \mathbf{R}_{\ell=1} = \delta\omega_{\mathrm{rot}}\,\mathrm{diag}(-1, 0, +1)
\end{equation}
where $\delta\omega_{\mathrm{rot}}\propto\Omega$.
Because $\mathbf{R}_\ell$ is purely diagonal (for any $\ell$), the off-diagonal elements of $\mathbf{R}_\ell+\mathbf{M}_\ell$ are totally set by magnetic effects.
The elements of the Lorentz matrix $\mathbf{M}_\ell$ are, in turn, highly dependent on the geometry of the magnetic field.
For illustrative purposes, we subsequently adopt a dipolar magnetic field misaligned with the rotation axis by an angle $\beta$, for which the $\ell=1$ Lorentz matrix is
\begin{equation} \label{Ml_matrix}
    \mathbf{M}_{\ell=1} = \frac{3}{20}\delta\omega_{\mathrm{mag}}
    \begin{pmatrix}
        7+C_\beta & -\sqrt{2}S_\beta & 1-C_\beta \\
        -\sqrt{2}S_\beta & 6-2C_\beta & \sqrt{2}S_\beta \\
        1-C_\beta & \sqrt{2}S_\beta & 7+C_\beta \\
    \end{pmatrix}
    \mathrm{,}
\end{equation}
where $C_\beta=\cos2\beta$, $S_\beta=\sin2\beta$, and $\delta\omega_{\mathrm{mag}}$ depends on some stellar interior-averaged magnetic field ($\delta\omega_{\mathrm{mag}}\propto\langle B_r^2\rangle$).
While this form of $\mathbf{M}_{\ell=1}$ only describes a specific magnetic field configuration, the qualitative results of this analysis generalize to other large-scale non-axisymmetric magnetic fields.

\new{Dipole mode frequency shifts can be decomposed into a mean component $\delta\bar{\nu}$ which shifts the entire triplet and an asymmetric component which changes the triplet's structure.}
Figure \ref{fig:figure_magnetic_shifts} shows the observed frequency structure of the rotational triplet as a function of $\delta\omega_{\mathrm{mag}}$\new{, for oblique g-modes with $\beta=70\degree$. Specifically, Figure \ref{fig:figure_magnetic_shifts} shows the rotational and magnetic frequency shifts with $\delta\bar{\nu}$ subtracted off, where $\delta\bar{\nu}$ is the mean frequency shift of the nine periodicities weighted by the observed spectral power $\propto|c_{j;\ell m}d^\ell_{0m}(i)|^2$.}
When the Coriolis force dominates ($\delta\omega_{\mathrm{mag}}\lesssim\delta\omega_{\mathrm{rot}}$; the \textit{blue region} in Figure \ref{fig:figure_magnetic_shifts}), the off-diagonal elements of $\mathbf{R}_\ell+\mathbf{M}_\ell$ are negligible, and magnetism shifts the mode frequencies without creating extra observed periodicities.
This is the familiar limit assumed by most observational \citep{li2022magnetic,deheuvels2023strong,li2023internal,hatt2024asteroseismic} and many theoretical \new{\citep{bugnet2021magnetic,bugnet2022magnetic,mathis2023asymmetries,das2024unveiling}} studies \new{\citep[although see][]{loi2021topology,li2022magnetic}}.
In this regime, magnetism causes rotational multiplets to be asymmetric, but does not introduce extra periodicities to the signal.

\begin{figure}
    \centering
    \includegraphics[width=0.48\textwidth]{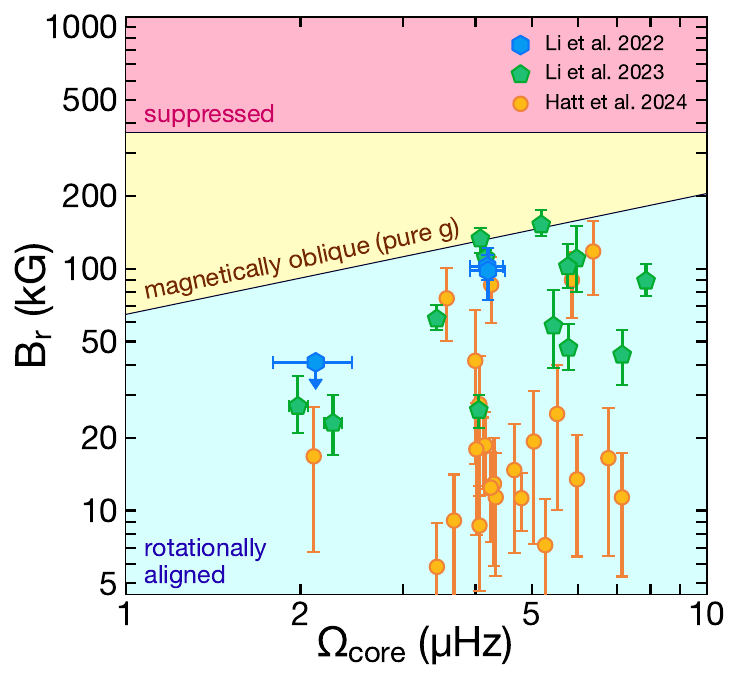}
    \caption{``Phase diagram'' of pulsation regimes in the space of magnetic field versus rotation rate, \new{where data points indicate known magnetic red giants with observed rotational triplets.}
    Regions of rotational alignment ($\delta\omega_{\mathrm{mag}}<\delta\omega_{\mathrm{rot}}$) and magnetic obliquity ($\delta\omega_{\mathrm{mag}}>\delta\omega_{\mathrm{rot}}$) are shown for the case of pure g modes\new{, assuming} the stellar model described in Section \ref{redgianttriplet} and the stellar parameters described in Section \ref{simulatedobservations}.
    The condition for magnetic g-mode suppression ($B>B_{\mathrm{crit}}$; Equation \ref{bcrit}) is also shown \new{for the same model}.
    \new{These shaded regions show approximate, model-dependent estimates for these regimes, and the precise boundaries between the regimes depend on stellar parameters and therefore vary from star to star.}
    }
    \label{fig:figure_magnetoalign_summary}
\end{figure}

In contrast, when the Lorentz force dominates ($\delta\omega_{\mathrm{mag}}\gtrsim\delta\omega_{\mathrm{rot}}$; the \textit{yellow region} in Figure \ref{fig:figure_magnetic_shifts}), the off-diagonal elements of $\mathbf{R}_\ell+\mathbf{M}_\ell$ cause the eigenvectors $\mathbf{c}_\ell$ to mix across $m$: the pulsations are oblique.
The rotationally aligned and magnetically oblique regimes are respectively indicated by the \textit{blue} and \textit{yellow regions} in Figure \ref{fig:figure_magnetoalign_summary} for a standard $M=1.5M_\odot$, $R=5.2R_\odot$ red giant stellar model generated using version r24.08.1 of Modules for Experiments in Stellar Astrophysics \citep[MESA;][]{paxton2010modules,paxton2013modules,paxton2015modules,paxton2018modules,paxton2019modules,jermyn2023modules}.
Magnetic fields detected thus far lie primarily in the rotationally aligned regime, \new{consistent with the modeling assumptions used to identify and interpret them.}
However, magnetically oblique pulsators should exist at slightly larger field strengths, especially for low core rotation rates \new{\citep[as also speculated by][]{li2023internal}}. 

There is a different (but overlapping) condition under which the magnetic field significantly alters the propagation of gravity waves, such that the weak-field theory breaks down.
This occurs for magnetic fields near a critical field strength which \citet{fuller2015asteroseismology} gives as
\begin{equation} \label{bcrit}
    B_{r,\mathrm{crit}} \simeq \sqrt{\frac{\pi}{\ell(\ell+1)}}\frac{\sqrt{\rho}\omega^2r}{N}\mathrm{.}
\end{equation}
Magnetic fields stronger than $B_{r,\mathrm{crit}}$ (the \textit{red region} in Figure \ref{fig:figure_magnetoalign_summary}) are expected to suppress g-mode propagation \citep{fuller2015asteroseismology,rui2023gravity,stello2016prevalence}.
\new{Notably, \citet{deheuvels2023strong} discovered and characterized $11$ red giants with distorted period spacing patterns consistent with relatively strong internal magnetic fields.
In particular, one of their stars (KIC 6975038) appears to experience magnetic suppression which exclusively affects its lowest-frequency mixed modes, consistent with theoretical expectations.}
The magnetically oblique and suppressed regions would shift to smaller field strengths for stars farther up the red giant branch. 

\subsection{Simulated observations of oblique pulsations} \label{simulatedobservations}

We simulate observations of a single g-mode triplet under the effects of a \new{strong inclined magnetic dipole} by numerically solving Equation \ref{MR_eigenvalueproblem} for the frequencies in the observer's inertial frame.
We choose parameters which produce the rotational triplet structure at the maximum value of \new{$\delta\nu_{\mathrm{mag}}=1.7\,\mu\mathrm{Hz}$} shown in Figure \ref{fig:figure_magnetic_shifts}.
In addition to the parameters described in Section \ref{redgianttriplet}, this corresponds for our red giant model to the realistic (though optimistic) rotation period $P_{\mathrm{rot}}=2\pi/\Omega=30\,\mathrm{d}$ and a field strength $B_r\approx292\,\mathrm{kG}$ (for a flat magnetic field \new{radial} profile).
We also assume a typical mode lifetime $\tau=2\,\mathrm{months}$.

\begin{figure*}
    \centering
    \includegraphics[width=\textwidth]{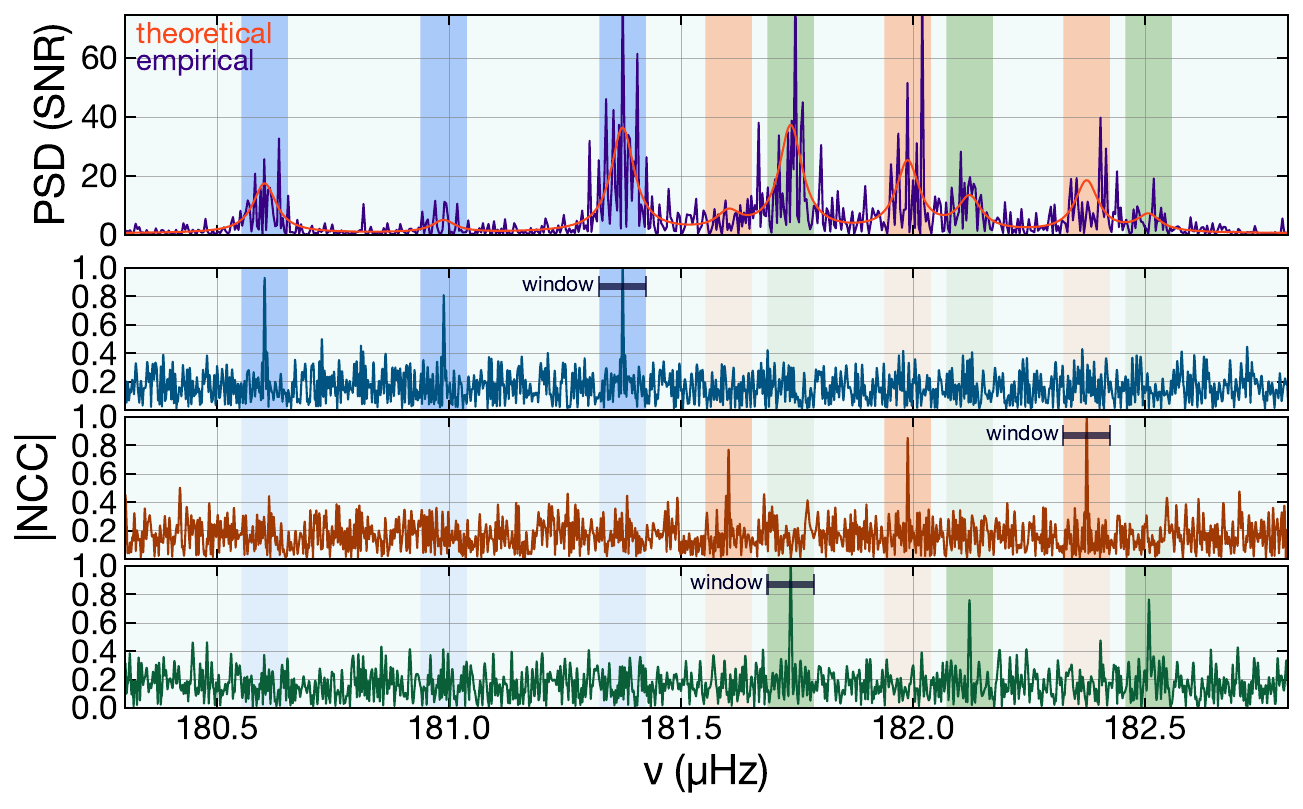}
    \caption{Synthetic data of a magnetically oblique g-mode triplet in a red giant, described further in Section \ref{redgianttriplet}).
    Background shading indicates the location of a periodicity, where the color represents the physical mode from which a periodicity originates.
    \textit{Top:} The observed PSD $|\delta\tilde{I}|^2$, normalized to the expectation value of the simulated white noise floor.
    \textit{Bottom:} The absolute value of the windowed normalized cross-correlation ($|\mathrm{NCC}|$) with respect to the three windows indicated by the black bars.
    $|\mathrm{NCC}|$ measures whether segments of the Fourier transform are spectrally correlated to the chosen window.
    As discussed in Section \ref{redgianttriplet}, $|\mathrm{NCC}|$ peaks at secondary periodicities whose corresponding mode matches the periodicity in the window, while mostly ignoring periodicities corresponding to different modes.
    }
    \label{fig:figure_clone_spectra}
\end{figure*}

Our mock observations are evenly sampled in time with a cadence $\Delta t=30\,\mathrm{min}$ and a baseline $T_{\mathrm{obs}}=8\,\mathrm{yr}$ \citep[twice that of the main \textit{Kepler} mission;][]{borucki2016kepler}.
We also inject a white noise background to mimic a realistic height-to-background ratio \citep[cf.][]{li2022magnetic}.
\new{Details of our procedure for generating synthetic data can be found in Appendix \ref{stochasticappendix}.}
The empirical PSD derived from these mock observations is shown in the \textit{top panel} of Figure \ref{fig:figure_clone_spectra}.
Figure \ref{fig:figure_clone_spectra} also shows the theoretically expected value of the PSD (\textit{orange curve}).
However, since the statistics of the process now explicitly modulate with time, the process is not stationary and is therefore not fully described by the PSD like in the usual case of a stationary process \citep[see, e.g.,][]{gardner2003measurement}.

We \new{perform a mock analysis of} our simulated observations by calculating a windowed, complex-valued version of a normalized cross correlation \citep[NCC;][]{kirch2008encyclopedia} of  $\delta\tilde{I}(\omega)$.
The NCC measures the overlap between two arrays, and is extensively applied to image comparison.
Specifically, we compute
\begin{equation} \label{ncc}
    \mathrm{NCC}_{\delta\omega_{\mathrm{win}}}(\omega_1,\omega_2) = \frac{\mathcal{I}_{\delta\omega_{\mathrm{win}}}(\omega_1,\omega_2)}{\sqrt{\mathcal{I}_{\delta\omega_{\mathrm{win}}}(\omega_1,\omega_1)\mathcal{I}_{\delta\omega_{\mathrm{win}}}(\omega_2,\omega_2)}}\mathrm{,}
\end{equation}
where we have defined the windowed inner product
\begin{equation}
    \mathcal{I}_{\delta\omega_{\mathrm{win}}}(\omega_1,\omega_2) \equiv \frac{1}{2\delta\omega_{\mathrm{win}}}\int^{+\delta\omega_{\mathrm{win}}}_{-\delta\omega_{\mathrm{win}}}\delta\tilde{I}^*(\omega_1+\xi)\delta\tilde{I}(\omega_2+\xi)\,\mathrm{d}\xi\mathrm{,}
\end{equation}
for some choice of window width $\delta\omega_{\mathrm{win}}$.
Intuitively, $\mathrm{NCC}_{\delta\omega_{\mathrm{win}}}(\omega_1,\omega_2)$ quantifies the degree of spectral correlation between two segments of the Fourier transform centered at $\omega_1$ and $\omega_2$ with widths $\delta\omega_{\mathrm{win}}$.
The absolute value of NCC is $1$ when the segments are exactly identical, up to a constant scaling factor, and tends to $0$ when the segments are totally spectrally uncorrelated.

In practice, owing to the finite baseline, we compute a discrete version of the windowed NCC by taking
\begin{equation} \label{discreteinnerprod}
    \mathcal{I}(\omega_1,\omega_2) \approx \sum^{+\delta n_{\mathrm{win}}}_{k=-\delta n_{\mathrm{win}}}\delta\tilde{I}_{n_{\omega_1}+k}^*\delta\tilde{I}_{n_{\omega_2}+k}\mathrm{,}
\end{equation}
where $\delta\tilde{I}_n$ denotes elements of the discrete Fourier transform, $\delta n_{\mathrm{win}}\approx\delta\omega_{\mathrm{win}}/\Delta\omega$ (where $\Delta\omega$ is the frequency resolution), and $n_{\omega_1}$ ($n_{\omega_2}$) refers to the index whose corresponding frequency is closest to $\omega_1$ ($\omega_2$).
To mitigate non-ideal frequency-resolution effects, we also sinc-interpolate the discrete Fourier transform by zero-padding the time series by a factor of ten \citep{schanze1995sinc}.

Each of the \textit{bottom panels} of Figure \ref{fig:figure_clone_spectra} shows the windowed NCC of our mock triplet, with $\omega_1$ fixed to the frequency of a different power spectral peak and the window width to
$\delta\omega_{\mathrm{win}}/2\pi=0.05\,\mu\mathrm{Hz}$.
When $\omega_1$ is chosen this way, the NCC measures how spectrally correlated segments of the Fourier transform are with the windowed peak.
It is thus expected to spike when $\omega_2$ coincides with the frequency of another periodicity arising from the same mode (i.e., another peak with the same shading in Figure \ref{fig:figure_clone_spectra}).
Indeed, Figure \ref{fig:figure_clone_spectra} shows that the NCC peaks substantially in absolute value at these other periodicities, which are often lower in amplitude (but still significant).
In contrast, the magnitude of the windowed NCC does not peak significantly at periodicities corresponding to other modes, since their different noise profiles destructively interfere with the window within the inner product in Equation \ref{discreteinnerprod}.
This proof-of-concept mock analysis demonstrates the ability to perform model-independent searches for oblique, stochastic pulsations in the frequency domain.

\section{Summary and prospects} \label{conclusion}

Sufficiently strong \new{non-axisymmetric} magnetic fields can misalign stellar pulsations from the rotation axis, i.e., cause the pulsations to be oblique.
\new{The oblique pulsator model is the standard framework for interpreting the pulsations of} rapidly oscillating Ap stars, which are known to harbor strong surface \new{oblique dipolar-like} magnetic fields \citep{kurtz1982rapidly,dziembowski1996magnetic}.
\new{Additionally, o}blique pulsations appear to have been detected in the prototype DBV white dwarf pulsator \citep[GD 358;][]{montgomery2010evidence} and have been suggested to produce the quintuplet of periodicities observed in the main-sequence pulsator $\beta$ Cephei \citep{telting1997period,shibahashi2000asteroseismology}.
\new{More recently, two blue large-amplitude pulsators (BLAPs) were also suggested to exhibit oblique pulsations \citep{PigulskiBLAPs}.}

In this work, we show that time series observations of oblique pulsators with stochasticity share generic signatures which can be used to identify them.
These signatures take the form of a permanent periodic mode amplitude modulation (in the time domain) or, equivalently, spectrally correlated line profiles in the Fourier transform (in the frequency domain).
Due to their dependence on stochasticity, these signatures are easiest to detect for modes with shorter lifetimes (i.e., larger linewidths).
Promisingly, searches for these signatures can be agnostic to the precise pattern of mode frequencies predicted by models.
Data analyses looking for spectrally correlated signals may identify magnetic red giants in the \textit{yellow region} in Figure \ref{fig:figure_magnetoalign_summary}, within which current analyses are inapplicable.

While magnetic red giants are the motivating use case for this work, our results apply to oblique pulsations in any pulsator with detectable stochasticity.
Outside of red giants, stochasticity is a prominent characteristic of main-sequence solar-like oscillators \citep[such as the Sun itself;][]{chaplin2013asteroseismology}, and has also been observed in some modes in classical pulsators \citep[such as $\delta$ Scuti stars;][]{breger_dsct1998} and compact pulsators \citep[such as white dwarfs and hot subdwarfs;][]{winget1994whole,hermes2017white,reed2007follow,ostensen2014stochastic}.

A similar analysis is likely also applicable to tidally tilted pulsations (TTPs), which are misaligned from the rotation axis by tidal forces from a companion rather than a strong magnetic field \citep{handler2020tidally,fuller2020tidally,fuller2025tidally}.
As such, TTPs are also oblique pulsations, and are similarly characterized by multiple periodicities with fixed relative phase and amplitude relationships \citep[e.g.,][]{fuller2025tidally}.
While TTPs have been discovered in coherent pulsators such as $\delta$ Scuti \citep{handler2020tidally} and subdwarf B \citep{jayaraman2022tidally} stars, searches for anomalous long-term coherence may enable the discovery of TTPs in stochastic pulsators.

Magnetic fields produce oblique pulsations when they are strong enough to overwhelm rotational effects such as the Coriolis force (i.e., for parameters within the \textit{yellow region} in Figure \ref{fig:figure_magnetoalign_summary}).
Magnetic fields which are close in strength to $B_{r,\mathrm{crit}}$ mix modes across values of $\ell$ \citep{lecoanet2017conversion,loi2020effect,dhouib2022detecting,rui2023gravity,rui2024asteroseismic,lecoanet2022asteroseismic}, not just $m$ as in this work.
This will produce additional peaks (e.g., more than three for $\ell=1$) spaced by the stellar spin frequency in the PSD, which can likely be detected by the same method outlined here.
In future work, we plan to extend our analysis of magnetic obliquity to stronger magnetic field strengths $B_r\sim B_{r,\mathrm{crit}}$.

${}$ 

\noindent We thank Emily Hatt for sharing a catalog of asteroseismic magnetic field measurements in a recent sample of red giants.
\new{We also thank Janosz Dewberry, Masao Takata, and the anonymous referee for their helpful comments.}
We are grateful for support from the United States--Israel Binational Science Foundation through grant BSF-2022175.
N.Z.R. acknowledges support from the National Science Foundation Graduate Research Fellowship under Grant No. DGE‐1745301.
J.M.J.O. acknowledges support from NASA through the NASA Hubble Fellowship grant HST-HF2-51517.001, awarded by STScI.
STScI is operated by the Association of Universities for Research in Astronomy, Incorporated, under NASA contract NAS5-26555.

\bibliography{bibliography.bib}

\appendix

\section{Transforming between the observer and corotating frames} \label{wignerappendix}

Performing the disk integral in Equation \ref{deltaI} requires a change of coordinates from the corotating frame (unprimed; $\theta,\phi$) to the observer's frame (primed; $\theta',\phi'$).
In this Appendix, we describe how spherical harmonics transform under this change of coordinates.

These two coordinate systems differ in relative rotation (by the stellar rotation rate $\Omega$) and choice of polar axis (by the inclination angle $i$).
The transformation between the two frames is characterized by the Euler angles $(\alpha_e,\beta_e,\gamma_e)=(0,i,\Omega t)$ \citep[in an intrinsic z-y-z convention, following][]{rose1995elementary}.
Under rotation, spherical harmonics transform as
\begin{equation} \label{wigner}
    Y_{\ell m}(\theta,\phi) = \sum_{m'}D^\ell_{m'm}(\alpha_e,\beta_e,\gamma_e)Y_{\ell m'}(\theta',\phi')\mathrm{,}
\end{equation}
where $D^\ell$ is the Wigner $D$ matrix, whose elements are
\begin{equation} \label{wignerD}
    D^\ell_{m'm}(\alpha_e,\beta_e,\gamma_e) = e^{-im'\alpha_e-im\gamma_e}d^\ell_{m'm}(\beta_e)\mathrm{,}
\end{equation}
where $d^\ell$ is a matrix which only depends on $\ell$.
For our particular transformation, 
\begin{equation} \label{wigner_specific_appendix}
    Y_{\ell m}(\theta,\phi) = \sum_{m'}e^{-im\Omega t}d^\ell_{m'm}(i)Y_{\ell m'}(\theta',\phi')\mathrm{,}
\end{equation}
reproducing Equation \ref{wigner_specific} in the main text.

\section{Amplitude \new{e}volution under \new{s}tochastic \new{d}riving} \label{stochasticappendix}

A stochastic mode is oscillatory (with a corotating frequency $\sigma_0$) on short timescales ($t\ll\tau$) but decoheres on longer timescales ($t\gg\tau$).
To mimic this behavior, we model the complex amplitude $A(t)$ as an underdamped harmonic oscillator with frequency \new{$\sigma_0$} and damping rate $\eta=1/\tau>0$ driven by noise $f(t)$.
The time dependence of the amplitudes follow
\begin{equation} \label{onesided_operator}
    \partial_tA(t) = (i\sigma_0-\eta)A(t) + \eta f(t)\mathrm{,}
\end{equation}
which is the stochastically driven amplitude equation \citep{buchler1993stellar} evaluated in the linear regime.
We have included an extra prefactor $\eta$ multiplying $f$ to normalize the height of the power spectral peak to be independent of $\eta$.
\new{Solutions to Equation \ref{onesided_operator} behave similarly to those of the usual damped driven harmonic oscillator equation, $\ddot{A}+2\eta\dot{A}+(\sigma_0^2+\eta^2)A=f$.
The difference is that the former excites only a single complex oscillation $\propto e^{i\sigma_0t}$ whereas the latter (as a second-order differential equation) excites linear combinations of two complex oscillations $\propto e^{i\sigma_0t}$ and $\propto e^{-i\sigma_0t}$.
Aiming to construct a simple excitation model for a single complex oscillation, we therefore solve Equation \ref{onesided_operator} rather than the usual damped driven harmonic oscillator equation.}

In the frequency domain, the Fourier transform of $A(t)$ is given by
\begin{equation} \label{AtildewrtFtilde}
    \tilde{A}(\sigma) = \frac{1}{1 + i(\sigma-\sigma_0)/\eta}\tilde{f}(\sigma)\mathrm{,}
\end{equation}
i.e., the product of a \new{peaked complex function whose squared norm is a Lorentzian and} the spectrum $\tilde{f}$ of the stochastic driving.
Although the square-root-Lorentzian factor has a complex phase profile (reflecting the phase lag of a damped driven harmonic oscillator), this phase is irrelevant as long as the stochastic driving is spectrally uncorrelated.

In the time domain, $A(t)$ is given by the inverse Fourier transform
\begin{equation} \label{At_appendix_orig}
    A(t) = \int^{+\infty}_{-\infty}\frac{\tilde{f}(\sigma)}{1+i(\sigma-\sigma_0)/\eta}e^{i\sigma t}\,\mathrm{d}\sigma\mathrm{.}
\end{equation}
Applying the change of variables $\sigma'=\sigma-\sigma_0$ allows us to rewrite Equation \ref{At_appendix_orig} as
\begin{equation} \label{At_appendix}
    A(t) = e^{i\sigma_0t}\int^{+\infty}_{-\infty}\frac{\tilde{f}(\sigma_0+\sigma')}{1+i\sigma'/\eta}e^{i\sigma't}\,\mathrm{d}\sigma'\mathrm{.}
\end{equation}
Defining
\begin{equation} \label{Abar}
    \bar{A}(t) \equiv \int^{+\infty}_{-\infty}\frac{\tilde{f}(\sigma_0+\sigma')}{1+i\sigma'/\eta}e^{i\sigma't}\,\mathrm{d}\sigma' \equiv \int^{+\infty}_{-\infty}\tilde{\bar{A}}(\sigma')e^{i\sigma't}\,\mathrm{d}\sigma'
\end{equation}
recovers Equation \ref{At_maintext} in the main text.
As a complex function, $\bar{A}(t)$ stores the departure in amplitude and phase of $A(t)$ from a perfect sinusoid.
Equation \ref{Abar} represents $\bar{A}(t)$ as the Fourier transform of a square-root Lorentzian multiplied by noise.
Since $\tilde{\bar{A}}$ is very roughly localized to frequencies $|\sigma'|\lesssim\eta$, $\bar{A}(t)$ remains roughly constant on short timescales $\lesssim\tau$ while varying randomly on longer timescales $\gg\tau$.

\new{For simplicity, we assume the driving to be white noise, although our results are not sensitive to this choice and extend to colored (but spectrally uncorrelated) noise.
To simulate a single statistical realization of $\tilde{A}(\sigma)$ on a discrete frequency grid (Section \ref{simulatedobservations}), we independently draw values from a $\chi^2$ distribution with two degrees of freedom, take their square roots, and multiply them by random complex phases drawn uniformly from the unit circle.
This ensures that the values of the PSD of the driving $|\tilde{f}(\sigma)|^2$  obey a $\chi^2$ distribution with two degrees of freedom, a standard assumption in the study of solar-like oscillations \citep{woodard1985short,duvall1986solar,anderson1990modeling} which results from the values of the real and imaginary parts of $\tilde{f}(\sigma)$ each individually following Gaussian distributions with identical statistics.
We then compute $\tilde{A}(\sigma)$ using Equation \ref{AtildewrtFtilde}.
The simulated time series $\delta I(t)$ is then obtained by applying an inverse fast Fourier transform to $\tilde{A}_j(\sigma)$ to obtain $A_j(t)$ for each mode $j$ and evaluating Equation \ref{combinedIfull}.
A simulated observational noise background is generated in the same way as $\tilde{f}(\sigma)$, and its inverse fast Fourier transform added to $\delta I(t)$.
}

\section{Rotating and magnetic g-mode frequencies and eigenfunctions} \label{magneticrg_appendix}

At lowest order, the simultaneous effects of rotation and magnetism can be calculated using degenerate perturbation theory.
In this Appendix, we quote the main results of this type of analysis for pure dipole ($\ell=1$) g modes \citep[in particular, those of][]{gomes2020core,bugnet2022magnetic,li2022magnetic,mathis2023asymmetries,das2024unveiling} in the asymptotic limit.
These conditions reasonably describe observable g modes in red giants.

The scale $\delta\omega_{\mathrm{rot}}$ of the Coriolis frequency shift is given by
\begin{equation} \label{deltaomegarot}
    \delta\omega_{\mathrm{rot}} = \frac{1}{2}\langle\Omega\rangle_g \equiv \frac{1}{2}\Omega\mathrm{,}
\end{equation}
where $\langle\Omega\rangle_g$ is the average rotation rate of the g-mode cavity (core).
Similarly, the scale of magnetic frequency shifts $\delta\omega_{\mathrm{mag}}$ is given by
\begin{equation} \label{deltaomega}
    \delta\omega_{\mathrm{mag}} = \frac{\mathscr{I}}{4\pi\omega_0^3}\langle B_r^2\rangle\mathrm{.}
\end{equation}
The value of $\delta\omega_{\mathrm{mag}}$ depends on a particular average of the squared radial component of the magnetic field over the g-mode cavity:
\begin{equation}
    \langle B_r^2\rangle = \int^{r_2}_{r_1}\mathrm{d}r\,K(r)\int_{S^2}\mathrm{d}\Omega_2\,B_r^2\mathrm{,}
\end{equation}
where $r_1$ and $r_2$ are the boundaries of the g-mode cavity (demarcated by $\omega^2<\min\left(S_1^2,N^2\right)$ where $S_1$ and $N$ are the $\ell=1$ Lamb and Brunt--V\"as\"al\"a frequencies) and $\mathrm{d}\Omega_2=\mathrm{d}(\cos\theta)\,\mathrm{d}\phi$ is an infinitesimal solid angle element.
This average is weighted in radius by a function
\begin{equation}
    K(r) \simeq
    \begin{cases}
        \frac{N^3/\rho r^3}{\int_{\mathcal{R}^\ell_{\;\nu}}(N^3/\rho r^3)\,\mathrm{d}r} & r_1\leq r\leq r_2 \\
        0 & \mathrm{otherwise,}
    \end{cases}
\end{equation}
which peaks near the hydrogen-burning shell in red giants. Finally, the sensitivity factor
\begin{equation}
    \mathscr{I} = \frac{\int^{r_2}_{r_1}(N^3/\rho r^3)\,\mathrm{d}r}{\int^{r_2}_{r_1}(N/r)\,\mathrm{d}r}
\end{equation} depends on the stellar structure.

For general magnetic configurations, the $\ell=1$ Lorentz force matrix $\mathbf{M}_{\ell=1}$ is
\begin{equation} \label{Mmatrixgeneral}
    \mathbf{M}_{\ell=1} = \delta\omega_{\mathrm{mag}}\frac{\int^{r_2}_{r_1}\mathrm{d}r\,K(r)\int_{S^2}\mathrm{d}\Omega_2\,\mathcal{M}_{\ell=1}(\theta,\phi)B_r^2}{\int^{r_2}_{r_1}\mathrm{d}r\,K(r)\int_{S^2}\mathrm{d}\Omega_2\,B_r^2}
\end{equation}
where
\begin{equation} \label{matrixweight}
    \mathbf{\mathcal{M}}_{\ell=1}(\theta,\phi) = \frac{3}{8}
    \begin{pmatrix}
        3+C_\theta & -\sqrt{2}e^{i\phi}S_\theta & e^{2i\phi}(1-C_\theta) \\
        -\sqrt{2}e^{-i\phi}S_\theta & 2-2C_\theta & \sqrt{2}e^{i\phi}S_\theta \\
        e^{-2i\phi}(1-C_\theta) & \sqrt{2}e^{-i\phi}S_\theta & 3+C_\theta \\
    \end{pmatrix}
\end{equation}
is a matrix-valued horizontal weighting function, with $S_\theta\equiv\sin2\theta$ and $C_\theta\equiv\cos2\theta$.
\new{Magnetic obliquity is generated by the off-diagonal elements of $\mathbf{M}_{\ell=1}$, which are related to the $m=1$ and $m=2$ coefficients of the azimuthal Fourier transform of $B_r^2$ \citep[see the Supplementary Information of][]{li2022magnetic}.
Azimuthal Fourier coefficients of higher $m$ correspond to smaller-scale azimuthal non-axisymmetry in the magnetic field which do not couple to the dipole modes in perturbation theory.}

In this work, we illustratively assume an inclined dipolar magnetic field such that $B_r\propto\cos\beta\cos\theta + \sin\beta\sin\theta\cos\phi$, where $\beta$ is the misalignment angle between the rotation and magnetic axes.
Combining Equations \ref{deltaomega}, \ref{Mmatrixgeneral}, and \ref{matrixweight} yields
\begin{equation} \label{Mmatrix_appendix}
    \mathbf{M}_{\ell=1} = \frac{3}{20}\delta\omega_{\mathrm{mag}}
    \begin{pmatrix}
        7+\cos2\beta & -\sqrt{2}\sin2\beta & 1-\cos2\beta \\
        -\sqrt{2}\sin2\beta & 6-2\cos2\beta & \sqrt{2}\sin2\beta \\
        1-\cos2\beta & \sqrt{2}\sin2\beta & 7+\cos2\beta \\
    \end{pmatrix}
    \mathrm{,}
\end{equation}
reproducing Equation \ref{Ml_matrix} in the text.

While this particular field geometry keeps the off-diagonal elements of $\mathbf{M}_{\ell=1}$ real, the off-diagonal elements can be complex in general.
Complex off-diagonal elements can cause phase shifts between periodicities associated with the same mode, but do not otherwise modify our general conclusions.

\end{document}